\documentclass[a4paper,5p,twocolumn]{elsarticle}
\usepackage{fancyhdr}
\usepackage{adjustbox}
\usepackage{multirow}
\usepackage{array}
\usepackage{hyperref}
\usepackage{booktabs}
\usepackage{amsmath}
\usepackage{amssymb}
\newcommand{\norm}[1]{\left\lVert#1\right\rVert}
\usepackage{float}
\usepackage[dvipsnames]{xcolor}
\usepackage{graphicx}      
\usepackage{natbib}        

\makeatletter
\def\endfigure{\end@float}
\def\endtable{\end@float}
\makeatother

\let\ifacconfcaptionwidth\captionwidth
\usepackage[caption=false]{subfig}
\let\captionwidth\ifacconfcaptionwidth

\begin{document}
\begin{frontmatter}

\title{Lane-Free Crossing of CAVs through Intersections as a Minimum-Time Optimal Control Problem} 


\author[First]{Mahdi Amouzadi} 
\author[First]{Mobolaji Olawumi Orisatoki} 
\author[First]{Arash M. Dizqah}

\address[First]{Smart Vehicles Control Laboratory (SVeCLab), University of Sussex, Brighton BN1 9RH, UK (e-mail: m.amouzadi@sussex.ac.uk).}

\begin{abstract}                
Unlike conventional cars, connected and autonomous vehicles (CAVs) can cross intersections in a lane-free order and utilise the whole area of intersections. This paper presents a minimum-time optimal control problem to centrally control the CAVs to simultaneously cross an intersection in the shortest possible time. Dual problem theory is employed to convexify the constraints of CAVs to avoid collision with each other and with road boundaries. The developed formulation is smooth and solvable by gradient-based algorithms. Simulation results show that the proposed strategy reduces the crossing time of intersections by an average of 52\% and 54\% as compared to, respectively, the state-of-the-art reservation-based and lane-free methods. Furthermore, the crossing time by the proposed strategy is fixed to a constant value for an intersection regardless of the number of CAVs.
\end{abstract}


\end{frontmatter}

\section{Introduction}

Traffic lights schedule the intersections inefficiently due to the limitations of human drivers, and this makes the intersections one of the major roots of congestion in urban traffic networks. CAVs are able to communicate with each other and with infrastructure to cross the intersections faster and safer, and hence to reduce energy consumption and increase traffic throughput \citep{rios2016survey}. However, such crossing of intersections is a complex problem and developing safe and efficient algorithms for this problem is still an open research topic. 

Previous studies proposed three approaches for CAVs to pass through intersections, including intersection-reservation, conflict-point-reservation and lane-free. The first approach reserves the whole intersection for one of the CAVs at a time. In this regard, \cite{pan2020optimal,zhang2017decentralized} designed algorithms based on solving an optimal control problem (OCP) to reserve the intersection for a period of time and avoid collisions. However, reservation of the whole intersection for one vehicle at a time is not very effective in terms of throughput. 

Conflict-point-reservation approach reserves a finite number of specific points (called conflict points) instead of the whole of the intersection. \cite{malikopoulos2018decentralized,mirheli2019consensus} designed intersection crossing algorithms using different number of conflict points for a four-leg intersection. The proposed algorithms enforce CAVs to reserve the approaching conflict point(s) prior to their arrival. A similar work is proposed in \cite{zhang2021priority}, where a constraint is added to the optimisation problem for each conflict point to limit the maximum number of crossing vehicles at any time to one. Although conflict-point-reservation approach improves the throughput of the intersection as compared to the intersection-reservation approach, yet vehicles are restricted to follow predefined paths and are not able to fully explore the intersection area. 

The lane-free approach allows CAVs to utilise the whole space of the intersection which can significantly improve the throughput of the intersection. Prior studies formulated the lane-free crossing of CAVs as an OCP \citep{li2020autonomous,li2018near}. To avoid collisions, the Euclidean distance between any pair of CAVs are constrained to be greater than a safe margin. This formulation of the collision avoidance constraints is non-convex \citep{colombo2012efficient}, and hence any optimisation problem including them is difficult to solve. \cite{li2018near} divided the non-convex problem of lane-free intersection crossing into two stages to make it tractable. At the first stage, the incoming vehicles are arranged as a specific form. Depending on the set of destinations of the formed CAVs, an offline solution is fetched from a lookup table. The approach is interesting, however, solving the offline non-convex problem for all possible crossing scenarios takes an impractical time. Alternatively, \cite{li2020autonomous} fixed the crossing time to a constant value and converted the minimum-time optimal control problem to a feasibility problem to solve online. However, finding a feasible (collision-free) solution to the problem of the intersection crossing does not fully exploit the CAVs' advantages to minimise the crossing time.

In summary, majority of the literature propose reservation-based algorithms which calculate the feasible collision-free crossing. To the best of the authors' knowledge, there is a limited number of literature dealing with lane-free minimum-time crossing of intersections which is a non-convex and computationally expensive problem. This paper addresses these gaps by formulating, convexifying and solving this problem. It shows that the resulting minimum crossing time is fixed to a constant value for a junction regardless of the number of CAVs. 

\begin{figure}[t]
    \centering
	\includegraphics[scale=0.45]{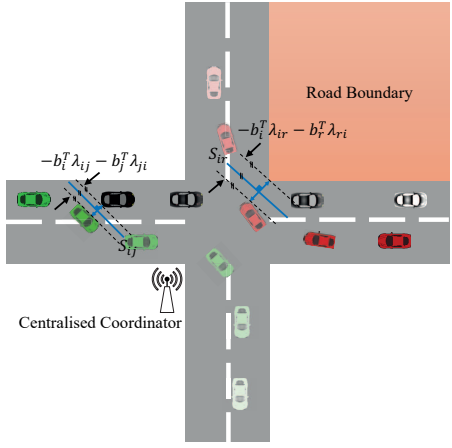}
	\caption{A lane-free and signal-free intersection.}
	\label{Intersection_layout}
\end{figure}

The remainder of this paper is organised as follows: Section \ref{system} describes the system of multiple CAVs crossing an intersection and presents the notations used throughout the paper; Section \ref{PF} formulates the lane-free crossing of CAVs through an intersection as a convexified minimum-time optimal control problem; Section \ref{Discussion} provides numerical results obtained from simulations along with discussions, and Section \ref{conclusion} concludes the outcomes.

\section{System Description}\label{system}
\subsection{Lane-Free and Signal-Free Intersections}
Fig. \ref{Intersection_layout} illustrates an example layout of a lane-free and signal-free intersection. The figure includes three CAVs which are moving from their initial points, depicted with the most solid colour, towards their intended destinations which are with the most transparent colour. The intersection comprises of four approaches, each of them has a separate incoming and outgoing lane. In a lane-free intersection, vehicles can freely change their lanes in favour of faster crossing through the intersection. For instance, Fig. \ref{Intersection_layout} shows the green CAV overtakes the black CAV by using the opposite lane. It is worth noting that the intersection does not have traffic lights because the CAVs can directly communicate their states and intentions. In this study, there is a coordinator that receives all the information from the CAVs and centrally control the vehicles to efficiently and safely cross the intersection. This fully-connected and centralised topology provides the optimal solution as the benchmark for testing any decentralised strategy.

The black and green CAVs also show the collision avoidance. The expression $-b_{i}^\top\lambda_{ij} - b_{j}^\top\lambda_{ji}$ is the dual representation of the distance between a pair of CAVs whilst $S_{ij}^*$ is the separating hyperplane placed between them. Similarly, $-b_{i}^\top\lambda_{ir} - b_{r}^\top\lambda_{ri}$ is the dual representation of distance between the red CAV and the coloured road boundary whilst $S_{ir}^*$ is the separating hyperplane between them. These equations are further explained in section \ref{PF}.  

\subsection{Modelling of CAV and road boundaries}
This study represents the lateral behaviour of a $\text{CAV}_{i}, \forall i\in \{1..N\}$ where $N$ is the number of crossing CAVs, with the bicycle model \citep{milliken1995race}. The bicycle model consists of two degree-of-freedoms (DoFs) which are sideslip angle $\beta_{i}$ and yaw rate $r_{i}$. The model also includes an additional DoF for the longitudinal velocity $V_{i}$. The pose of the $\text{CAV}_{i}$ at each time $t$ with respect to its initial pose is then calculated from these states. The equations of the DoFs along with the pose of $\text{CAV}_{i}$ construct a set of non-linear differential equations as follows:

\vspace{-10pt}
\begin{align}
    \frac{d}{dt} \begin{bmatrix}\notag
    r_{i}\\[0.1cm]
    \beta_{i}\\[0.1cm]
    V_{i}\\[0.1cm]
    x_{i}\\[0.1cm]
    y_{i}\\[0.1cm]
    \theta_{i}\\[0.1cm]
    \end{bmatrix}(t)
    =
    &\begin{bmatrix}
    \frac{\tilde{N}_{r}}{I_{z}V_(t)} \cdot r_{i}(t) + \frac{N_{\beta}}{I_{z}} \cdot \beta_{i}(t)\\[0.1cm]
    (\frac{\tilde{Y}_{r}}{m \cdot V_{i}^2(t)} - 1) \cdot r_{i}(t) + \frac{Y_{\beta}}{m\cdot V_{i}(t)} \cdot \beta_{i}(t) \\[0.1cm]
     0\\[0.1cm]
    V_{i}(t) \cdot cos\theta_{i}(t)\\[0.1cm]
    V_{i}(t) \cdot sin\theta_{i}(t)\\[0.1cm]
    r_{i}(t)\\[0.1cm]
    \end{bmatrix}
    +\\ 
    &\begin{bmatrix}
    0 & \frac{N_{\delta}}{I_{z}} \\ 
    0 & \frac{Y_{\delta}}{m \cdot V_{i}(t)}  \\ 
    1 & 0 \\
    0 & 0 \\
    0 & 0 \\ 
    0 & 0 \\ 
    \end{bmatrix}
    \begin{bmatrix}
    a_{i}\\
    \delta_{i}
    \end{bmatrix}(t)
    ,
    t \in [t_{0},t_{f}].
    \label{vehilce_model}
\end{align}
where $\textbf{x} = [r_{i},\beta_{i},V_{i},x_{i},y_{i},\theta_{i}]^T$ and $\textbf{u} =[a_{i}, \delta_{i}]^T$ are, respectively, the system states and control inputs of $\text{CAV}_i$. $a_{i}(t)$ and $\delta_{i}(t)$ are, respectively, the acceleration ($m/s^2$) and steering angle ($rad$) of the vehicle. $\textbf{z}_{i}=[x_{i},y_{i},\theta_{i}]^T$ refers to the pose of $\text{CAV}_i$ in non-inertial reference system. The constants $m$ and $I_{z}$ denote mass ($kg$) and moment of inertia ($kg.m^2$) of the vehicle around axis $z$, respectively. $t_{0}$ and $t_{f}$ ($s$) represent the starting and ending time of the crossing. The calculation of vehicle parameters $\tilde{N}_r$, $N_\beta$, $N_\delta$, $\tilde{Y}_r$, $Y_\beta$ and $Y_\delta$ is shown in \citep{milliken1995race}.

To ensure CAVs drive within their dynamic limitations, the following constraints are enforced for each $\text{CAV}_i$:

\vspace{-10pt}
\begin{subequations}\label{Limits}
    \begin{align}
        \underline{V} \leq V_{i}(t)\leq \bar{V},\\
        \underline{a} \leq \rvert a_{i}(t) \rvert \leq \bar{a},\\
        \underline{\delta} \leq \rvert \delta_{i}(t) \rvert \leq \bar{\delta},\\
        \underline{r} \leq \rvert r_{i}(t) \rvert \leq \bar{r},\\
        \underline{\beta} \leq \rvert \beta_{i}(t) \rvert \leq \bar{\beta}.
     \end{align}
\end{subequations}
where $\overline{.}$ and $\underbar{.}$ are, respectively, the upper and lower boundaries.

Each $\text{CAV}_{i}$ is presented as a rectangular polytope $\zeta_{i}$ at origin where as the intersection of half-space linear inequality $A_{i}X \le b_{i}$ where $X = [x,y]^T$ is a Cartesian point. 

Road boundaries are also modelled with convex polytopic sets $O_{r}$ where $r \in \{1..N_{r}\}$ and $N_{r}=4$ denotes the total number of road boundaries for a four-legged intersection.

\section{Control Problem Formulation}\label{PF}
This section presents a minimum-time OCP that minimises the crossing time of an intersection whilst avoiding collisions of CAVs with others and with road boundaries.

\subsection{Constraints to Avoid Collisions Between CAVs}\label{CAVs_CAC}
To avoid collisions between any $\text{CAV}_i$ and $\text{CAV}_j$ $\forall i\neq j \in \{1..N\}$, their polytopic sets should not intersect, i.e. $\zeta_{i} \cap \zeta_{j} = \emptyset$ where $\zeta_{i} = \{\text{X} \in \mathbb{R}^{2}|A_{i}\text{X} \leq b_{i}\}$ and $\zeta_{j} = \{\text{Y} \in \mathbb{R}^{2}|A_{j}\text{Y} \leq b_{j}\}$. However, this is a non-convex and non-differentiable constraint. To enforce differentiability, $\zeta_{i} \cap \zeta_{j} = \emptyset$ is replaced by the following sufficient condition \citep{zhang2020optimization}: 

\vspace{-10pt}
\begin{align}\label{primal}
    dist (\zeta_{i},\zeta_{j}) =\underset{\text{X,Y}}{\text{min}}\{\norm{\text{X-Y}}_{2}\;|\;A_{i}\text{X}&\leq b_{i},\; A_{j}\text{Y}\leq b_{j}\} \geq d_{min};\notag \\ 
    &\forall i\neq j \in \{1..N\}.
\end{align}
where $d_{min}$ is a minimum safety distance between any pair of CAVs.

It is known that the problem of finding the minimum distance between a polytope $\zeta_{i}$ and another given polytope $\zeta_{j}$ is convex \citep{boyd_vandenberghe_2004}. Therefore the substituting sufficient condition, $dist (\zeta_{i},\zeta_{j}) \geq d_{min}$, is convex and since $\zeta_{j}$ is not an empty set, the strong duality holds \citep{zhang2020optimization}. This means that the solution of the primal problem $dist (\zeta_{i},\zeta_{j}) \geq d_{min}$ is the same as the one of its dual problem which is as follows:

\vspace{-10pt}
\begin{align}\label{dual}
  \text {dist} (\zeta_{i},\zeta_{j}) :=\:&\underset{\lambda_{ij},\lambda_{ji},s_{ij}}{\text{max}} -b_{i}^\top\lambda_{ij} - b_{j}^\top\lambda_{ji} \\
&\text{s.t.} \notag \quad A_{i}^\top\lambda_{ij} + s_{ij} = 0, A_{j}^\top\lambda_{ji} - s_{ij} = 0, \notag\\ 
&\quad \quad  \norm{s_{ij}}_{2} \leq 1, -\lambda_{ij} \leq 0, -\lambda_{ji} \leq 0; \notag\\
& \quad \quad  \forall i\neq j \in \{1..N\}.\notag
\end{align}
where $\lambda_{ij}$, $\lambda_{ji}$, and $s_{ij}$ are the dual variables. $A_{i}$ and $b_{i}$ represent the $\text{CAV}_{i}$'s polytope at each time step $t$, and are functions of the $\text{CAV}_i$'s pose $\textbf{z}_{i}(t)$. The derivation of dual problem (\ref{dual}) from primal problem (\ref{primal}) is shown in \cite{firoozi2020distributed}. 

Combining (\ref{dual}) with (\ref{primal}), (\ref{dual}) can be substituted by $\{\exists \lambda_{ij} \geq 0, \lambda_{ji} \geq 0, s_{ij} : -b_{i}^\top\lambda_{ij} - b_{j}^\top\lambda_{ji} \geq d_{min}, A_{i}^\top\lambda_{ij} + s_{ij} = 0, A_{j}^\top\lambda_{ji} - s_{ij} = 0, \norm{s_{ij}}_{2} \leq 1\}$ because the existence of a feasible solution $\lambda_{ij,feas}$, $\lambda_{ji,feas}$, and $s_{ij,feas}$ where $-b_{i}^\top\lambda_{ij,feas} - b_{j}^\top\lambda_{ji,feas} \geq d_{min}$ is a sufficient condition to ensure $ dist (\zeta_{i},\zeta_{j}) \geq d_{min}$, i.e. to avoid collisions \citep{firoozi2020distributed}. As seen in Fig. \ref{Intersection_layout}, $S_{ij}$ is a separating hyperplane between $\text{CAV}_i$ and $\text{CAV}_j$, and $S_{ij}$ = $S_{ji}$.

\subsection{Constraints to Avoid Collisions with Road Boundaries}\label{CAVs_ROAD}
Each $\text{CAV}_{i}$ must also avoid all the road boundaries, i.e. $\zeta_{i} \cap O_{r} = \emptyset$ where $\zeta_{i} = \{\text{X} \in \mathbb{R}^{2}|A_{i}\text{X} \leq b_{i}\}$ and $O_{r} =\{\text{Y} \in \mathbb{R}^{2}|A_{r}\text{Y} \leq b_{r}\}$. Similar to section \ref{CAVs_CAC}, $\zeta_{i} \cap O_{r} = \emptyset$ is replaced by the following sufficient condition:

\vspace{-10pt}
\begin{align}\label{primal_rd}
    \text {dist} (\zeta_{i},O_{r}) =\underset{\text{X,Y}}{\text{min}}\{\norm{\text{X-Y}}_{2}|A_{i}\text{X}\leq &b_{i},A_{r}\text{Y}\leq b_{r}\} \geq d_{rmin}; \notag \\
    &\forall r \in \{1..N_{r}\}.
\end{align}
where $d_{rmin}$ is the minimum safety distance between CAVs and road boundaries.

The dual problem of (\ref{primal_rd}) is then substituted with the sufficient condition $\{\exists \lambda_{ir} \geq 0, \lambda_{ri} \geq 0, s_{ir} : -b_{i}^\top\lambda_{ir} - b_{r}^\top\lambda_{ri} \geq d_{rmin}, A_{i}^\top\lambda_{ir} + s_{ir} = 0, A_{j}^\top\lambda_{ri} - s_{ir} = 0, \norm{s_{ir}}_{2} \leq 1\}$
where $\lambda_{ir},\lambda_{ri}$, and $s_{ir}$ are the dual variables. $S_{ir}$ is the separating hyperplane between CAVs and road boundaries (see Fig. \ref{Intersection_layout}). 

\subsection{Objective Function}
CAVs are expected to reach their terminal pose as fast as possible. Therefore, this paper proposes the following minimum-time objective function that also penalises deviations between the real and intended final poses:
\begin{align}\label{cost}
   & J(\textbf{z}_1(.),..,\textbf{z}_N(.)) =  \alpha(t_{f} - t_{0})^2\nonumber + \\
    &\int_{t_{0}}^{t_{f}} \sum_{i=1}^{N}\,[(\textbf{z}_{i}(t)-\textbf{z}_{i}(t_{f}))^\top \textbf{Q} (\textbf{z}_{i}(t)- \textbf{z}_i(t_{f}))]\: dt \:
\end{align}
\\
where $\alpha$ and $\textbf{Q}$ are the gain factors which are selected based on trial and error. The expression $(t_{f}-t_{0})^2$ penalises the crossing time which is time that all the CAVs have crossed the intersection. The Lagrange term penalises deviations of the current pose $\textbf{z}_{i}(t)$ from the final pose $\textbf{z}_{i}(t_{f})$. The final pose of CAVs $\textbf{z}_{i}(t_{f})$ is chosen randomly and indicates the intended destination of each $\text{CAV}_{i}$. It is worth noting that the final poses are just for crossing the intersection where the algorithm is applied and CAVs can continue to their journey after passing the intersection.
 
\subsection{Optimal Control Problem}

Putting all together, lane-free crossing of $N$ CAVs through a signal-free intersection is formulated as the following smooth OCP:

\vspace{-10pt}
\begin{subequations} \label{OCP}
\begin{alignat}{10}
&\{t_{f},a_i(.),\delta_i(.)\}^* = \\ &\quad \quad\text{arg}\:\underset{\substack{t_{f},a_i(.),\delta_i(.)\\\lambda_{ij},\lambda_{ji},s_{ij},\\\lambda_{ri},\lambda_{ir},s_{ir}}}{\text{min}}  J(\textbf{z}_1(.),..,\textbf{z}_N(.)) := (\ref{cost}), \notag \\
 & \quad \quad \text{s.t.} \quad (\ref{vehilce_model}), (\ref{Limits}), \label{DSA}\\
 &\qquad \qquad    -b_{i}(\textbf{z}_{i}(t))^\top\lambda_{ij}(t)-b_{j}(\textbf{z}_{j}(t))^\top \lambda_{ji}(t)\ge d_{min} \label{CA_A} \\
 &\qquad \qquad   A_{i}(\textbf{z}_{i}(t))^\top \lambda_{ij}(t)+s_{ij}(t)=0 \label{CA_B}\\
 &\qquad \qquad    A_{j}(\textbf{z}_{j}(t))^\top \lambda_{ji}(t)-s_{ij}(t)=0 \label{CA_C} \\
  & \qquad \qquad   -b_{i}(\textbf{z}_{i}(t))^\top\lambda_{ir}(t)-b_{r}^\top \lambda_{ri}(t)\ge d_{rmin} \label{RA_A}\\
 &\qquad \qquad    A_{i}(\textbf{z}_{i}(t))^\top \lambda_{ir}(t)+s_{ir}(t)=0 \label{RA_B}\\
 &\qquad \qquad    A_{r}^\top \lambda_{ri}(t)+s_{ir}(t)=0 \label{RA_C}\\
 &\qquad \qquad    \lambda_{ij}(t),\;\lambda_{ji}(t),\;\lambda_{ir}(t),\;\lambda_{ri}(t)\geq 0, \\ 
 & \qquad \qquad   \norm{s_{ij}(t)}_{2}\leq 1,\;\norm{s_{ir}(t)}_{2}\leq1,\\
 &  \qquad \qquad  \textbf{z}_i(t_{0}) = \textbf{z}_{i,0},\\
 &  \qquad \qquad   \forall i\neq j \in \{1..N\}, \forall r \in \{1..N_r\}. \notag
\end{alignat}
\end{subequations}
where (\ref{DSA}) refers to the vehicle kinematics and CAVs' limitations. (\ref{CA_A}, \ref{CA_B} and \ref{CA_C}) and (\ref{RA_A}, \ref{RA_B} and \ref{RA_C}) denote, respectively, the constraints of CAVs to avoid collisions with each other and with road boundaries.

Solution of the problem (\ref{OCP}) is the shortest possible terminal time $t_{f}$, as well as the optimal trajectories of the control signals $a_i(.)^*$ and $\delta_i(.)^*$ of each $\text{CAV}_i$ over $\forall t\in[t_0,t_f]$. The initial pose $\textbf{z}_i(t_{0})$ and initial speed $V_i$ of all $\text{CAV}_i \: \forall i \in \{1..N\}$ within the control zone are known. The remaining of the states are also assumed as zero. These initial conditions at $t = t_{0}$ are feasible solutions of the OCP. 

\section{Numerical Simulation and Discussion}\label{Discussion}

This paper employs CasADi \citep{Andersson2019} and IPOPT \citep{wachter2006implementation} to solve the formulated nonlinear OCP (\ref{OCP}). CasAdi directly discretises the continuous-time OCP (\ref{OCP}), using a collocation method, to construct an equivalent nonlinear programming (NLP) \citep{rosmann2020time}. The resulting NLP is then solved using interior-point method (IPM) \citep{wachter2006implementation}. The paper also improves computation time by linking IPOPT to Intel® oneAPI Math Kernel Library (oneMKL, \color{blue} https://software.intel.com\color{black}), which includes high-performance implementation of the \textit{MA27} linear solver.

This section verifies performance of the developed control strategy by running two test scenarios with different number of CAVs. All the results are calculated with Matlab running on a Linux Ubuntu server with a 3.7 GHz Intel core i7 and 32 GB memory. Table \ref{settings} provides the chosen values for important parameters of the algorithm. In order to assist with understanding, a video that visualises the results of the two test scenarios is also provided on \color{blue}\url{https://youtu.be/L_aFGkKT38U}\color{black}.

\subsection{Test Scenario One: Comparison to the state-of-the-art}

The first test scenario compares the effectiveness of the proposed algorithm in terms of crossing time, average and standard deviation of speeds against two state-of-the-art methods. The scenario consists of testing for different number of CAVs between 2 and 12. All tests consist of right-turn turning maneuver by at least one CAV. The starting and terminal positions, permissible maximum speed and acceleration of CAVs are the same for all the cases and are shown in table \ref{settings}.  

The first benchmark is the conflict-point-reservation approach in \cite{malikopoulos2021optimal}, where each CAV calculates its own trajectory by jointly minimising the travelling time and energy consumption and stores its reservation time for each conflict point in a coordinator. The second benchmark is the lane-free method proposed by \cite{li2020autonomous}. This work calculates the control inputs for a fixed final time, however, there is no reservation of conflict points and the CAVs can freely use all the space of the junction, as long as there is no collision.

\renewcommand{\arraystretch}{1}
\begin{table}[t]
\centering
    \caption{Main parameters of the model} 
    \label{settings} 
    \begin{tabular}{l l c}
    \toprule
    \textbf{Parameter(s)} & \textbf{Description}  & \textbf{Value(s)}\\ [0.5ex] 
    \midrule
    $\text{d}_{min}\: (m)$ & \multicolumn{1}{m{3.7cm}}{minimum distance between CAVs} & 0.1  \\
    $\text{d}_{rmin} (m)$ & \multicolumn{1}{m{3.7cm}}{minimum distance between CAVs-road boundaries} & 0.1 \\
    $d$ (-) & \multicolumn{1}{m{3.7cm}}{number of collocation points}  & 5  \\[0.5ex]
    $N_{p}\:$(-) & \multicolumn{1}{m{3.7cm}}{prediction horizon (prediction step is calculated)}  & 15  \\[0.5ex]
    $V_{max}\: (m/s)$ & \multicolumn{1}{m{3.7cm}}{bounds on $V_{i}$} & 25  \\[0.5ex]
    $\delta_{max}\:(rad)$ & \multicolumn{1}{m{3.7cm}}{bounds on $|\delta_{i}|$} & 0.67 \\[0.5ex]
    $a_{max}\:(m/s^2)$ & \multicolumn{1}{m{3.7cm}}{bounds on $|a_{i}|$} & 3 \\[0.5ex]
    $r_{i}$ (m) & \multicolumn{1}{m{3.7cm}}{bounds on $|r_{i}|$} & 0.7 \\[0.5ex]
    $\beta_{max}$ (rad/s) & \multicolumn{1}{m{3.7cm}}{bounds on $|\beta_{i}|$} & 0.5 \\[0.5ex]
    \multicolumn{1}{m{2.5cm}}{$V_i(t_0)\; \forall i \in \{1..N\}$ $\,(m/s)$}& initial speed & 10 \\[0.5ex]
    \bottomrule
    \end{tabular}
\end{table}

\begin{table}[t]
\centering
    \caption{A comparison between the proposed method and two state-of-the-art for test scenario one.}
    \begin{tabular}{l m{0.46cm} m{0.46cm} m{0.46cm} m{0.46cm} m{0.46cm} m{0.46cm} m{0.46cm}}
    \toprule
    \textbf{Number of CAVs} & \textbf{2} & \textbf{4} & \textbf{6} & \textbf{8} & \textbf{10} & \textbf{12} \\ \hline
    \multicolumn{1}{m{3.1cm}}{\textbf{The proposed work}} & & & & & & \\
    \multicolumn{1}{m{3.1cm}}{Min. crossing time (s)}  & 4.6 & 4.6 & 4.6   & 4.6   & 4.6   & 4.6\\ 
    \multicolumn{1}{m{3.1cm}}{Average speed (m/s)}  & 15.2 & 15.8 & 15.6   & 15.1   & 15.4   & 15.5\\ 
    \multicolumn{1}{m{3.1cm}}{Standard deviation of speed}  & 3.46 & 3.77 & 3.66   & 3.43   & 3.58   & 3.66\\\hline
    \multicolumn{1}{m{3.1cm}}{\textbf{\cite{malikopoulos2021optimal}}} & & & & & & \\    
    \multicolumn{1}{m{3.1cm}}{Crossing time (s)}  & 6.3 & 6.3 & 11.3   & 12.9   & 12.9  & 12.9\\ 
    \multicolumn{1}{m{3.1cm}}{Average speed (m/s)}  & 12.7 & 13.5 & 11.3   & 9.9   & 10.6   & 11.1 \\ 
    \multicolumn{1}{m{3.1cm}}{Standard deviation of speed }  & 2.3 & 2.5 & 4.4   & 4.4   & 4.6   & 4.6\\\hline
    \multicolumn{1}{m{3.1cm}}{\textbf{\cite{li2020autonomous}}} & & & & & & \\
    \multicolumn{1}{m{3.1cm}}{Crossing time (s) }  & 10 & 10 & 10   & 10   & 10   & 10\\ 
    \multicolumn{1}{m{3.1cm}}{Average speed (m/s)}  & 10.1 & 10.0 & 10.0   & 10.0   & 10.1   & 10.0\\ 
    \multicolumn{1}{m{3.1cm}}{Standard deviation of speed}  & 0.1 & 0.1 & 0.1   & 0.2   & 0.2   & 0.2\\
    \bottomrule
    \end{tabular}\label{benchmark_results}
\end{table}

The results in Table \ref{benchmark_results} show that the proposed strategy improves the crossing time of CAVs by, respectively, up to $65\%$ (and an average improvement of $52\%$ for different number of CAVs), and $54\%$ as compared to the reservation-based approach in \cite{malikopoulos2021optimal} and lane-free method in \cite{li2020autonomous}. Moreover, the average speed of CAVs by the proposed algorithm is larger (for larger number of crossing CAVs) and the standard deviation of speeds are smaller than the ones by the reservation-based method. Also, these values by the proposed work are almost fixed to constant values regardless of the number of crossing CAVs as opposed to the reservation-based method. These suggest that by using the proposed algorithm CAVs are crossing the intersection quicker and also more consistently. The reservation-based methods restrain the motion of CAVs to orderly pass the critical points which results in a higher crossing time and lower average speed values than the proposed strategy. Furthermore, the lane-free method in \cite{li2020autonomous} fixes the crossing time to make the calculation tractable, however, it lowers average speeds and therefore increases crossing time. 

\begin{figure}[t]
    \centering
    \includegraphics[scale=0.3]{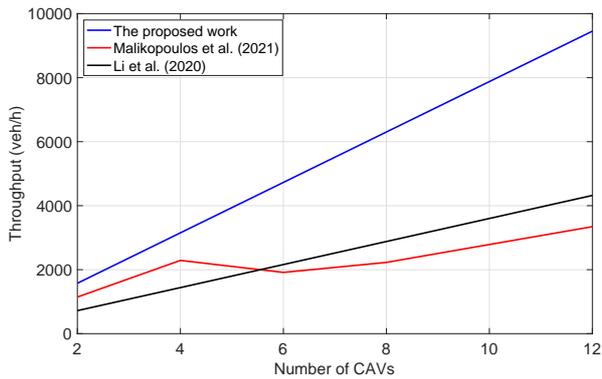}
    \caption{Throughput at different number of CAVs for test scenario one.}
    \label{throughput}
\end{figure}

Fig. \ref{throughput} shows that the resulting throughput of the proposed algorithm is higher than the ones of both the reservation-based and lane-free methods. Throughput of the lane-free intersection linearly increases in terms of number of CAVs as opposed to the reservation-based approach. Analytical determination of the maximum throughput (i.e., capacity) is a future work.

\subsection{Test Scenario Two: A more complex scenario}
The second test evaluates performance of the proposed strategy for a complex scenario that involves 21 CAVs and allows any direction of travel (right, left, and straight).

Table \ref{complex_Ta} compares the crossing time, average and standard deviation of speeds for different number of CAVs for test scenario two. As seen in Table \ref{complex_Ta}, the crossing time by the proposed algorithm does not vary by changing the test scenario and again is fixed to a constant value regardless of the number of CAVs. This is an interesting outcome showing the optimal crossing time of lane-free junctions is limited by the layout of the junction rather than the number of passing CAVs as in traditional signalised junctions. 

The shortest time to cross the intersection is achieved when the CAV that travels the longest distance amongst all the crossing CAVs can cross the intersection with the maximum permissible acceleration. In both test scenarios CAVs travel up to 70 meters that means the shortest crossing time with the maximum acceleration of $3\ (m/s^2)$ and initial speed of $10\ (m/s)$ is 4.27 $s$. The results show that the proposed algorithm can be as close as desired to this value in cost of deviation from the final point.

\begin{table}[t]
\centering
    \caption{Simulation results of test scenario two for different number of CAVs.}
    \begin{tabular}{l m{0.26cm} m{0.26cm} m{0.26cm} m{0.26cm} m{0.26cm} m{0.26cm} m{0.26cm}}
    \toprule
    \textbf{Number of CAVs} & \textbf{3}      & \textbf{6}      & \textbf{9} & \textbf{12} & \textbf{15} & \textbf{18} & \textbf{21}  \\ \hline
    Min. crossing time (s)  & 4.5 & 4.6 & 4.6   & 4.6   & 4.6   & 4.6 & 4.6\\
    Average speed (m/s)  & 13.2 & 14.5 & 14.5   & 14.2   & 14.1   & 13.8  &  13.6\\
    Standard deviation of speed  & 3.9 & 4.0 & 3.8 & 3.9 & 4.1 & 4.2 & 4.1\\
    \bottomrule
    \end{tabular}\label{complex_Ta}
\end{table}
\begin{figure}[t]
    \centering
    \includegraphics[scale=0.45]{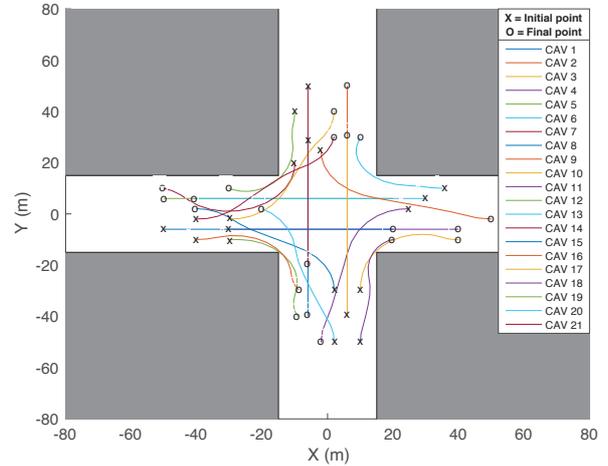}
    \caption{The optimal trajectories which are calculated by the proposed strategy for 21 CAVs crossing a junction according to the test scenario two.}
    \label{Trajectory}
\end{figure}

\begin{figure}[t]
    \centering
    \includegraphics[scale=0.45]{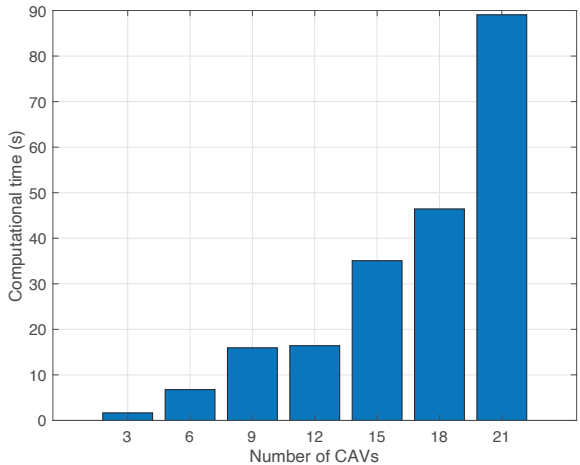}
    \caption{Means of the computational time of 10 runs of the proposed strategy for different number of CAVs in test scenario two. The standard deviation of the 10 runs for each number of CAVs is less than 0.5\%.}
    \label{ComTime}
\end{figure}

Fig. \ref{Trajectory} illustrates the calculated motion trajectories of all CAVs in the second test scenario. As seen, CAVs freely move and use opposite lanes if it is collision-free while avoiding road boundaries. Please check out a provided video on \color{blue}\url{https://youtu.be/L_aFGkKT38U} \color{black} for more details. 

\subsection{Computational Time}
The computational time of test scenario two for each number of crossing CAVs is presented in Fig. \ref{ComTime}. The presented numbers are the average of 10 times of running the scenario. The standard deviation of the results is less than $0.5\%$ and is not shown. Fig. \ref{ComTime} shows that the computational complexity of the proposed algorithm is of the order of $O(e^{0.13N})$ in terms of number of crossing CAVs $N$. The future work of this study is to reduce the computational time by decentralising the proposed strategy. 

\section{Conclusion}\label{conclusion}

This study proposes a convexified optimal control problem to minimise the crossing time of CAVs passing through intersections in a lane-free order. The simulation results show that the proposed strategy significantly reduces the crossing time and hence increases traffic throughput, as compared to the state-of-the-art reservation-based and lane-free strategies. Also, the results show that the crossing time by the developed controller only relies on the layout of the intersection and is independent of the number or manoeuvres of the crossing CAVs. In other words, the throughput of the intersection increases linearly with the number of crossing CAVs. The authors believe that the proposed algorithm provides a benchmark to evaluate the performance of other centralised or decentralised strategies in terms of throughput.


\bibliographystyle{elsarticle-harv}
\bibliography{ifacconf.bib}             

\begin{thebibliography}{17}
\expandafter\ifx\csname natexlab\endcsname\relax\def\natexlab#1{#1}\fi
\providecommand{\url}[1]{\texttt{#1}}
\providecommand{\href}[2]{#2}
\providecommand{\path}[1]{#1}
\providecommand{\DOIprefix}{doi:}
\providecommand{\ArXivprefix}{arXiv:}
\providecommand{\URLprefix}{URL: }
\providecommand{\Pubmedprefix}{pmid:}
\providecommand{\doi}[1]{\href{http://dx.doi.org/#1}{\path{#1}}}
\providecommand{\Pubmed}[1]{\href{pmid:#1}{\path{#1}}}
\providecommand{\bibinfo}[2]{#2}
\ifx\xfnm\relax \def\xfnm[#1]{\unskip,\space#1}\fi
\bibitem[{Andersson et~al.(2019)Andersson, Gillis, Horn, Rawlings and
  Diehl}]{Andersson2019}
\bibinfo{author}{Andersson, J.A.E.}, \bibinfo{author}{Gillis, J.},
  \bibinfo{author}{Horn, G.}, \bibinfo{author}{Rawlings, J.B.},
  \bibinfo{author}{Diehl, M.}, \bibinfo{year}{2019}.
\newblock \bibinfo{title}{{CasADi} -- {A} software framework for nonlinear
  optimization and optimal control}.
\newblock \bibinfo{journal}{Mathematical Programming Computation}
  \bibinfo{volume}{11}, \bibinfo{pages}{1--36}.
\newblock \DOIprefix\doi{10.1007/s12532-018-0139-4}.
\bibitem[{Boyd and Vandenberghe(2004)}]{boyd_vandenberghe_2004}
\bibinfo{author}{Boyd, S.}, \bibinfo{author}{Vandenberghe, L.},
  \bibinfo{year}{2004}.
\newblock \bibinfo{title}{Convex Optimization}.
\newblock \bibinfo{publisher}{Cambridge University Press}.
\newblock \DOIprefix\doi{10.1017/CBO9780511804441}.
\bibitem[{Colombo and Del~Vecchio(2012)}]{colombo2012efficient}
\bibinfo{author}{Colombo, A.}, \bibinfo{author}{Del~Vecchio, D.},
  \bibinfo{year}{2012}.
\newblock \bibinfo{title}{Efficient algorithms for collision avoidance at
  intersections}, in: \bibinfo{booktitle}{Proceedings of the 15th ACM
  international conference on Hybrid Systems: Computation and Control}, pp.
  \bibinfo{pages}{145--154}.
\bibitem[{Firoozi et~al.(2020)Firoozi, Ferranti, Zhang, Nejadnik and
  Borrelli}]{firoozi2020distributed}
\bibinfo{author}{Firoozi, R.}, \bibinfo{author}{Ferranti, L.},
  \bibinfo{author}{Zhang, X.}, \bibinfo{author}{Nejadnik, S.},
  \bibinfo{author}{Borrelli, F.}, \bibinfo{year}{2020}.
\newblock \bibinfo{title}{A distributed multi-robot coordination algorithm for
  navigation in tight environments}.
\newblock \bibinfo{journal}{arXiv preprint arXiv:2006.11492} .
\bibitem[{Li et~al.(2020)Li, Zhang, Jia and Peng}]{li2020autonomous}
\bibinfo{author}{Li, B.}, \bibinfo{author}{Zhang, Y.}, \bibinfo{author}{Jia,
  N.}, \bibinfo{author}{Peng, X.}, \bibinfo{year}{2020}.
\newblock \bibinfo{title}{Autonomous intersection management over continuous
  space: A microscopic and precise solution via computational optimal control}.
\newblock \bibinfo{journal}{IFAC-PapersOnLine} \bibinfo{volume}{53},
  \bibinfo{pages}{17071--17076}.
\bibitem[{Li et~al.(2018)Li, Zhang, Zhang, Jia and Ge}]{li2018near}
\bibinfo{author}{Li, B.}, \bibinfo{author}{Zhang, Y.}, \bibinfo{author}{Zhang,
  Y.}, \bibinfo{author}{Jia, N.}, \bibinfo{author}{Ge, Y.},
  \bibinfo{year}{2018}.
\newblock \bibinfo{title}{Near-optimal online motion planning of connected and
  automated vehicles at a signal-free and lane-free intersection}, in:
  \bibinfo{booktitle}{2018 IEEE Intelligent Vehicles Symposium (IV)},
  \bibinfo{organization}{IEEE}. pp. \bibinfo{pages}{1432--1437}.
\bibitem[{Malikopoulos et~al.(2021)Malikopoulos, Beaver and
  Chremos}]{malikopoulos2021optimal}
\bibinfo{author}{Malikopoulos, A.A.}, \bibinfo{author}{Beaver, L.},
  \bibinfo{author}{Chremos, I.V.}, \bibinfo{year}{2021}.
\newblock \bibinfo{title}{Optimal time trajectory and coordination for
  connected and automated vehicles}.
\newblock \bibinfo{journal}{Automatica} \bibinfo{volume}{125},
  \bibinfo{pages}{109469}.
\bibitem[{Malikopoulos et~al.(2018)Malikopoulos, Cassandras and
  Zhang}]{malikopoulos2018decentralized}
\bibinfo{author}{Malikopoulos, A.A.}, \bibinfo{author}{Cassandras, C.G.},
  \bibinfo{author}{Zhang, Y.J.}, \bibinfo{year}{2018}.
\newblock \bibinfo{title}{A decentralized energy-optimal control framework for
  connected automated vehicles at signal-free intersections}.
\newblock \bibinfo{journal}{Automatica} \bibinfo{volume}{93},
  \bibinfo{pages}{244--256}.
\bibitem[{Milliken and Milliken(1995)}]{milliken1995race}
\bibinfo{author}{Milliken, W.}, \bibinfo{author}{Milliken, D.},
  \bibinfo{year}{1995}.
\newblock \bibinfo{title}{Race Car Vehicle Dynamics}.
\newblock Premiere Series, \bibinfo{publisher}{SAE International}.
\newblock \URLprefix \url{https://books.google.co.uk/books?id=opgHfQzlnLEC}.
\bibitem[{Mirheli et~al.(2019)Mirheli, Tajalli, Hajibabai and
  Hajbabaie}]{mirheli2019consensus}
\bibinfo{author}{Mirheli, A.}, \bibinfo{author}{Tajalli, M.},
  \bibinfo{author}{Hajibabai, L.}, \bibinfo{author}{Hajbabaie, A.},
  \bibinfo{year}{2019}.
\newblock \bibinfo{title}{A consensus-based distributed trajectory control in a
  signal-free intersection}.
\newblock \bibinfo{journal}{Transportation research part C: emerging
  technologies} \bibinfo{volume}{100}, \bibinfo{pages}{161--176}.
\bibitem[{Pan et~al.(2020)Pan, Chen, Evangelou and Timotheou}]{pan2020optimal}
\bibinfo{author}{Pan, X.}, \bibinfo{author}{Chen, B.},
  \bibinfo{author}{Evangelou, S.A.}, \bibinfo{author}{Timotheou, S.},
  \bibinfo{year}{2020}.
\newblock \bibinfo{title}{Optimal motion control for connected and automated
  electric vehicles at signal-free intersections}, in: \bibinfo{booktitle}{2020
  59th IEEE Conference on Decision and Control (CDC)},
  \bibinfo{organization}{IEEE}. pp. \bibinfo{pages}{2831--2836}.
\bibitem[{Rios-Torres and Malikopoulos(2016)}]{rios2016survey}
\bibinfo{author}{Rios-Torres, J.}, \bibinfo{author}{Malikopoulos, A.A.},
  \bibinfo{year}{2016}.
\newblock \bibinfo{title}{A survey on the coordination of connected and
  automated vehicles at intersections and merging at highway on-ramps}.
\newblock \bibinfo{journal}{IEEE Transactions on Intelligent Transportation
  Systems} \bibinfo{volume}{18}, \bibinfo{pages}{1066--1077}.
\bibitem[{R{\"o}smann et~al.(2020)R{\"o}smann, Makarow and
  Bertram}]{rosmann2020time}
\bibinfo{author}{R{\"o}smann, C.}, \bibinfo{author}{Makarow, A.},
  \bibinfo{author}{Bertram, T.}, \bibinfo{year}{2020}.
\newblock \bibinfo{title}{Time-optimal control with direct collocation and
  variable discretization}.
\newblock \bibinfo{journal}{arXiv preprint arXiv:2005.12136} .
\bibitem[{W{\"a}chter and Biegler(2006)}]{wachter2006implementation}
\bibinfo{author}{W{\"a}chter, A.}, \bibinfo{author}{Biegler, L.T.},
  \bibinfo{year}{2006}.
\newblock \bibinfo{title}{On the implementation of an interior-point filter
  line-search algorithm for large-scale nonlinear programming}.
\newblock \bibinfo{journal}{Mathematical programming} \bibinfo{volume}{106},
  \bibinfo{pages}{25--57}.
\bibitem[{Zhang et~al.(2021a)Zhang, Zhang, Chen, Duan, Cheng and
  Yang}]{zhang2021priority}
\bibinfo{author}{Zhang, H.}, \bibinfo{author}{Zhang, R.},
  \bibinfo{author}{Chen, C.}, \bibinfo{author}{Duan, D.},
  \bibinfo{author}{Cheng, X.}, \bibinfo{author}{Yang, L.},
  \bibinfo{year}{2021}a.
\newblock \bibinfo{title}{A priority-based autonomous intersection management
  (aim) scheme for connected automated vehicles (cavs)}.
\newblock \bibinfo{journal}{Vehicles} \bibinfo{volume}{3},
  \bibinfo{pages}{533--544}.
\bibitem[{Zhang et~al.(2021b)Zhang, Liniger and
  Borrelli}]{zhang2020optimization}
\bibinfo{author}{Zhang, X.}, \bibinfo{author}{Liniger, A.},
  \bibinfo{author}{Borrelli, F.}, \bibinfo{year}{2021}b.
\newblock \bibinfo{title}{Optimization-based collision avoidance}.
\newblock \bibinfo{journal}{IEEE Transactions on Control Systems Technology}
  \bibinfo{volume}{29}, \bibinfo{pages}{972--983}.
\bibitem[{Zhang et~al.(2017)Zhang, Malikopoulos and
  Cassandras}]{zhang2017decentralized}
\bibinfo{author}{Zhang, Y.}, \bibinfo{author}{Malikopoulos, A.A.},
  \bibinfo{author}{Cassandras, C.G.}, \bibinfo{year}{2017}.
\newblock \bibinfo{title}{Decentralized optimal control for connected automated
  vehicles at intersections including left and right turns}, in:
  \bibinfo{booktitle}{2017 IEEE 56th Annual Conference on Decision and Control
  (CDC)}, \bibinfo{organization}{IEEE}. pp. \bibinfo{pages}{4428--4433}.

\end{thebibliography}
                                                   







\end{document}